\documentclass[pre,twocolumn,showpacs,floatfix,superscriptaddress,aps]{revtex4}
\usepackage{graphicx}
\usepackage{amsmath,amssymb}

\begin{document}

\title{Multifractal regime transition in a modified minority game model}

\author{Antonio  F. Crepaldi}

\affiliation{Departamento de Engenharia de Produ\c{c}\~ao,
Universidade Estadual Paulista, Av. Eng. Luiz Edmundo C. Coube 14-01
17033-360   , 01405-900 Bauru-SP, Brazil}

\author{Camilo Rodrigues Neto}
\author{Fernando F. Ferreira}

\affiliation{Grupo Interdisciplinar de F\'{\i}sica da
Informa\c{c}\~ao e Economia (GRIFE), Escola de Arte, Ci\^encias e
Humanidades, Universidade de S\~ao Paulo,  Av. Arlindo Bettio 1000,
03828-000 S\~ao Paulo, Brazil}

\author{Gerson Francisco}

\affiliation{Instituto de F\'{\i}sica Te\'orica, Universidade
Estadual Paulista, R.  Pamplona 145, 01405-900 S\~ao Paulo, Brazil}


\begin{abstract}

The search for more realistic modeling of financial time series
reveals several stylized facts of real markets. In this work we
focus on the multifractal properties found in price and index
signals. Although the usual Minority Game (MG) models do not exhibit
multifractality, we study here one of its variants that does. We
show that the nonsynchronous MG models in the nonergodic phase is
multifractal and in this sense, together with other stylized facts,
constitute a better modeling tool. Using the Structure Function (SF)
approach we detected the stationary and the scaling range of the
time series generated by the MG model and, from the linear
(nonlinear) behavior of the SF we identified the fractal
(multifractal) regimes. Finally, using the Wavelet Transform Modulus
Maxima (WTMM) technique we obtained its multifractal spectrum width
for different dynamical regimes.

\end{abstract}

\maketitle

\section{Introduction}

In the last years several agents based models for asset returns have
been proposed in the literature \cite{lux99,levylevy}. One of the
reasons for this interest is that the traditional models derived
from the geometrical Brownian motion do not explain adequately many
properties of real markets. The Minority Game (MG) and its variants
constitute one of  the most promising models \cite{Challet} due to
their capability of explaining a wider range of properties found in
price and index signals. It is usual to characterize economic time
series from their empirical properties called stylized facts
\cite{rama2001}, which includes multifractal long range
correlations. There are several ways to characterize the long-range
correlations from the real time series and from its models. Some of
these methods are the autocorrelation functions, power spectral
densities (either from Fourier or wavelets transforms) and
probability distribution functions. In addition, the fractal and the
multifractal analysis provide more insights, respectively, on the self-similar
and self-affine scaling exponents.
Here we use the Hurst exponent obtained from the
Structure Function (SF) method \cite{sf}, and the singularity
spectrum obtained from the Wavelet Transform Modulus Maxima (WTMM)
method \cite{muzy91, arneodo95} to determine, respectively, the
fractal and the multifractal structure of signals generated by the
Minority Game model.

The MG models are an oversimplified version of complex systems.
Their peculiarity is that several modifications on the original
model are possible in order to make it more realistic and
analytically tractable in the stationary state via replica or
generating functional methods \cite{Challet98, Challet99, fgeratriz,
Coolen}. Such methods have led to a deep understanding of their
macroscopic properties. Unfortunately, many interesting phenomena
occur in the nonergodic phase where the analytical approach fails.
Here we address a version of MG where the synchronicity of time
transactions is removed \cite{andersen2006, matteo2002, andrea2003}
allowing agents to trade on different time-scales. Remarkably, for
different distribution of frequencies, it has been shown that these
models essentially preserve many of the statistical features of
synchronized MG versions such as phase transition, volatility
clustering, fat tail PDFs and so on. However, these features now
depend on the distribution of the trading frequency. In the present
work we will focus on the case where the amount of information
processed by agents is the same.

This investigation is part of our continuing quest to find models
that comply with all known stylized facts in financial time
evolution \cite{fff2003}. Thus it is important to explore the
property of multiscaling in the artificially generated data. Many
empirical studies have indicated the presence of multifractal
behavior in real data as a stylized fact. Surprisingly, MG models
incorporate this effect through a more realistic trading rules of
agents.

The paper is organized as follows. In Section 2 we describe the
model, and then we discuss in Section 3 the statistical properties
of the return time series. In Section 4 we explain the multifractal
concept and the method used to detect it. The results of the
multifractal analysis are discussed in the Section 5. Finally, in
the last section we present our conclusions.

\section{Minority Game with different time-scales }
\label{}

The simplest version of the Minority Game \cite{Challet97} consists
on a set of $N_{s}$ adaptive agents, also called speculators. They
are endowed with $S$ strategies that map public information $\mu \in
[{0,\dots,P-1}]$ to the decision of buying or selling assets,
$a^{\mu}_{i,s}(t)=\pm 1$, $i=1\ldots N_{s}$, $s=1\ldots S$. Each
strategy is generated at the beginning of the game and is kept
frozen along the dynamics. Besides, the strategies have a payoff
function, $U_{i,s}(t)$, that describes their performance all the
time according to the following payoff function:
\begin{equation}
  U_{i,s}(t+1)=U_{i,s}(t)-a_{i,s}^{\mu}(t)A(t),
\end{equation}
where $A(t)=\sum_{i}a^{\mu}_{i,s^{*}}(t)$ is the excess demand. This
rule is applied to all strategies independently of their previous
scores. The public information is a non negative integer number
drawn at random each round uniformly from in the interval $[0,P)$.
The decisions are made according to $a^{\mu}_{i,s^{*}}(t)$, where
$s^{*}$ is the label for the best ranked strategy $(s^{*}=
max_{s}~U_{i,s})$. In case of a tie, the decision is made based on
coin tossing.

An improved model is the grand canonical minority game which
consists of two distinct groups, namely, producers and speculators.
The producers have one strategy, their decisions are a function of
$\mu$ only, and hence they are always doing transactions in the
market. The speculators now have an extra strategy, $s=0$, that
allows them not to trade when their strategies are not profitable
enough. We compute the strategies' performance with the following
payoff function
\begin{equation}
 U_{i,s}(t+1)=U_{i,s}(t)-a_{i,s}^{\mu}(t)A(t)+\epsilon \delta_{0,s}
\end{equation}
where $\epsilon$ is a small real number,
which can be positive (interest rate) or negative (agent's risk aversion measure).
Thanks to the null strategy, the number of
traders fluctuate along the time evolution.

Agents in the standard MG models trade on the same time-scales and
different events also occur with the same frequency. We can improve
this simplified assumption by introducing $g$ groups of $N_s$
speculators, each one with different time scale $ts_{j}$, where one
group $g_{j}$ play a singular minority game or grand canonical
minority game for $j=1\ldots g$. All agents, independently of the
group, have access to the same information $\mu$ and their payoff
function are updated virtually all the time. We will implement the
grand canonical version, with only one producer's group with $N_{p}$
members taking action in the smallest timescale. Those timescales
are chosen to be daily, weekly and monthly to resemble the real
market. The order parameters take into account the total number of
agents and it is written as $\alpha=\frac{P}{gN}$.

Other authors \cite{andersen2006, matteo2002, andrea2003} have
studied similar models. The basic picture of the behavior of the MG,
even considering different distribution of trading time-scales,
remains analogous to the synchronized trading frequency model. The
control parameter $\alpha$ exhibits some dependence with the
distribution frequency parameter.
When the time-scale difference
$|ts_{i}-ts_{j}|$ increases, for $i \ne j$, the groups become
independent and behave as a monochromatic MG. Curiously, the market
has the largest fluctuations for small time-scale difference, i.e.,
when the groups process data that are very close in time.

\section{Statistical analysis}
\label{stati}

The minority game models and their variants were extensively
explored in several studies. Their capability to generate universal
statistical properties qualitatively similar to the financial market
have been the  main reason for this interest. From the statistical
point of view, the known fact is that the probability distribution
of returns in the minority game time series is Gaussian except
around a critical point. In the subcritical region it is possible to
find certain realizations with fat tails but not necessarily
following a power law decay. When one allows interaction between
groups composed by the players  trading in different time scales,
the statistical properties are qualitatively preserved.

\begin{figure}[!ht]
  \centering
  \includegraphics[width=\linewidth,clip]{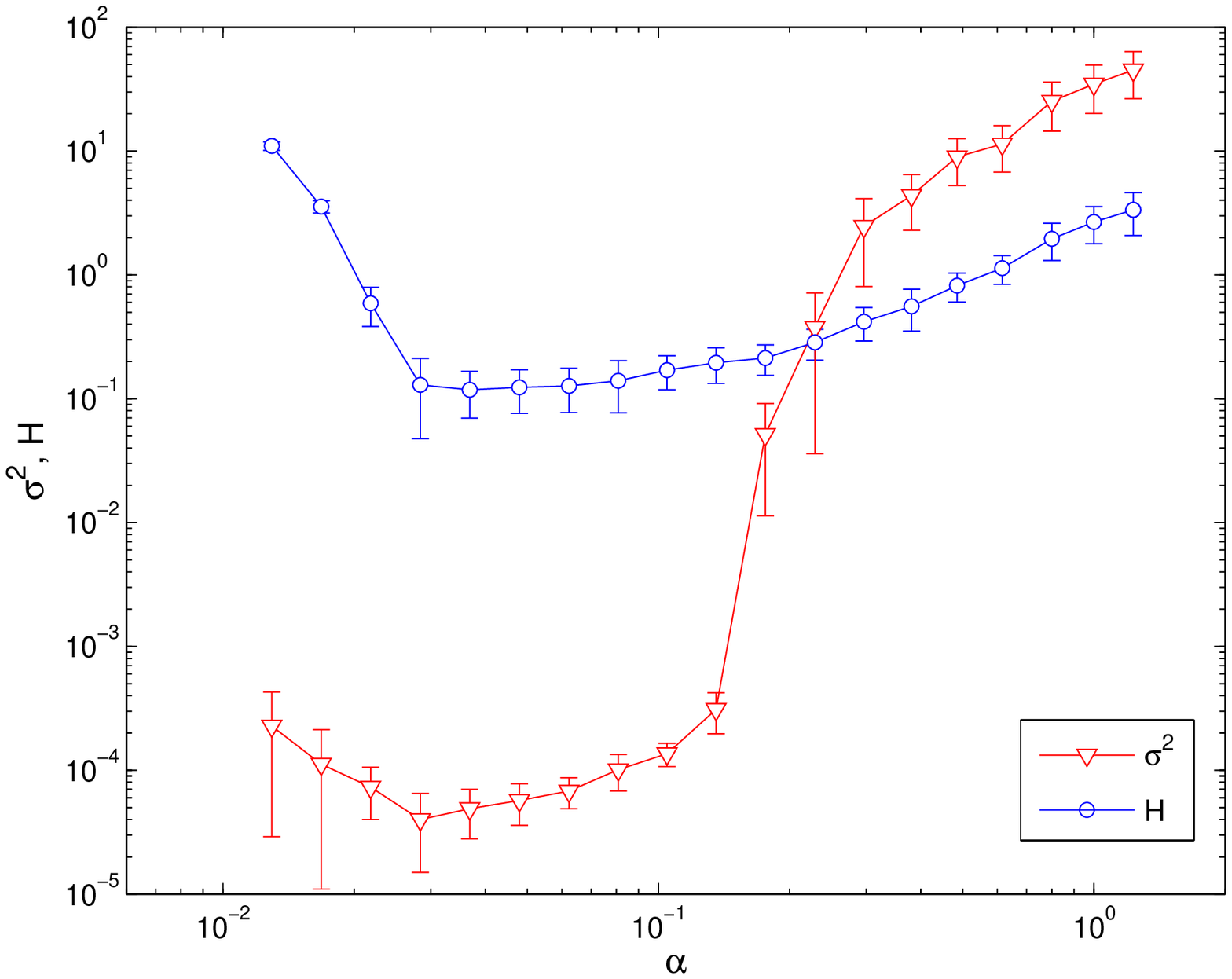}
  \includegraphics[width=\linewidth,clip]{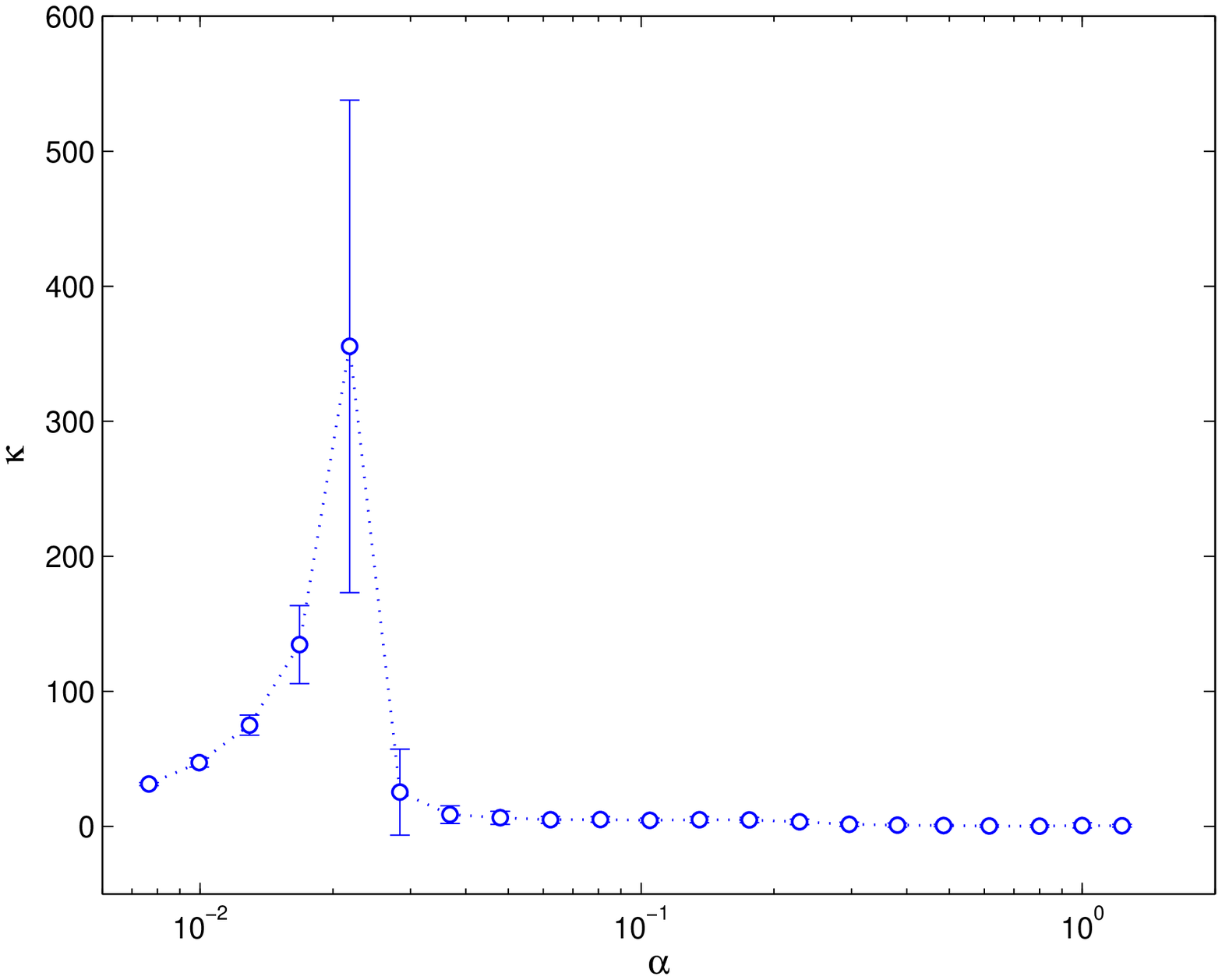}
  \caption{Top: phase diagram for $\sigma$ and H function.
    Bottom: Kurtosis.
    The critical value of $\alpha$ occurs for the minimum of the $\sigma$ curve, around $\alpha_c=0.2$.
    The error bars were obtained from 40 realizations of different initial
    conditions.} \label{kurt}
\end{figure}
Fig. \ref{kurt}(a) presents the variance $\sigma^2$ and the
predictability $H$ (as a function of $\alpha$) of the return $A(t)$
defined as:
\begin{eqnarray}
H= \frac{1}{P}\sum_{\mu=0}^{P-1} \langle A|\mu \rangle^2
\end{eqnarray}
where $\langle\ldots \rangle$ is the temporal average. From this
plot we can see the phase transition. The critical $\alpha_c$
separates the ergodic phase from the nonergodic phase. For $\alpha>
\alpha_c$ the time series generated by this model has a normal
distribution. Fig.~\ref{kurt}(a) shows the plots of $\sigma^2$
(upper plot) and $H$ (lower plot) against $\alpha$. Qualitatively
these these diagrams agree with the synchronous MG model and also
they help to locate the value of $\alpha_c$. Using the criterion
that $\alpha_c$ is the minimum of $\sigma$ we find that this value
is near 0.2. In Fig.~\ref{kurt}(b) we exhibit the kurtosis, denoted
by $\kappa$ also as a function of $\alpha$. Firstly,  in the low
$\alpha$ region, many realizations have high kurtosis and large
error bars around $\alpha=0.02$, caused by eventual coordination
among agents. It can be shown that for small values of $\alpha$ the
time series is leptokurtic.

The next measure we looked at were the probability density functions (PDF)
of the fluctuations at different scales. These are simply histograms of the height differences
\begin{equation}
\label{eq:pdf} \Delta Y_{\tau} = Y(t+\tau) - Y(t),
\end{equation}
where $Y(t)=\sum_{{t^\prime}=0}^t A(t^\prime)$ and $\tau$ is the scale. We will also refer to $\tau$ as the scale of analysis.

\begin{figure}[!ht]
 \centering
 \includegraphics[width=\linewidth,clip]{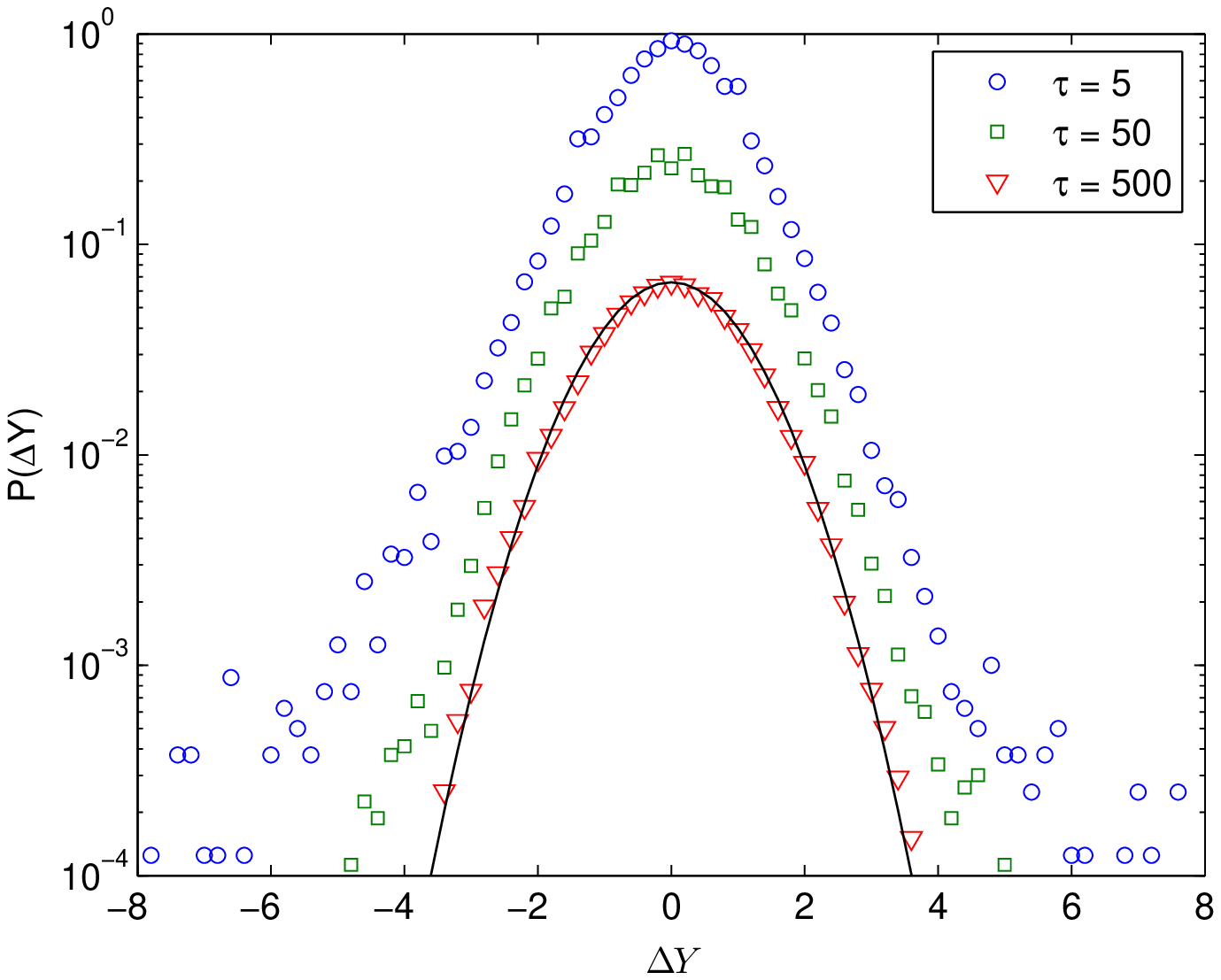}
 \includegraphics[width=\linewidth,clip]{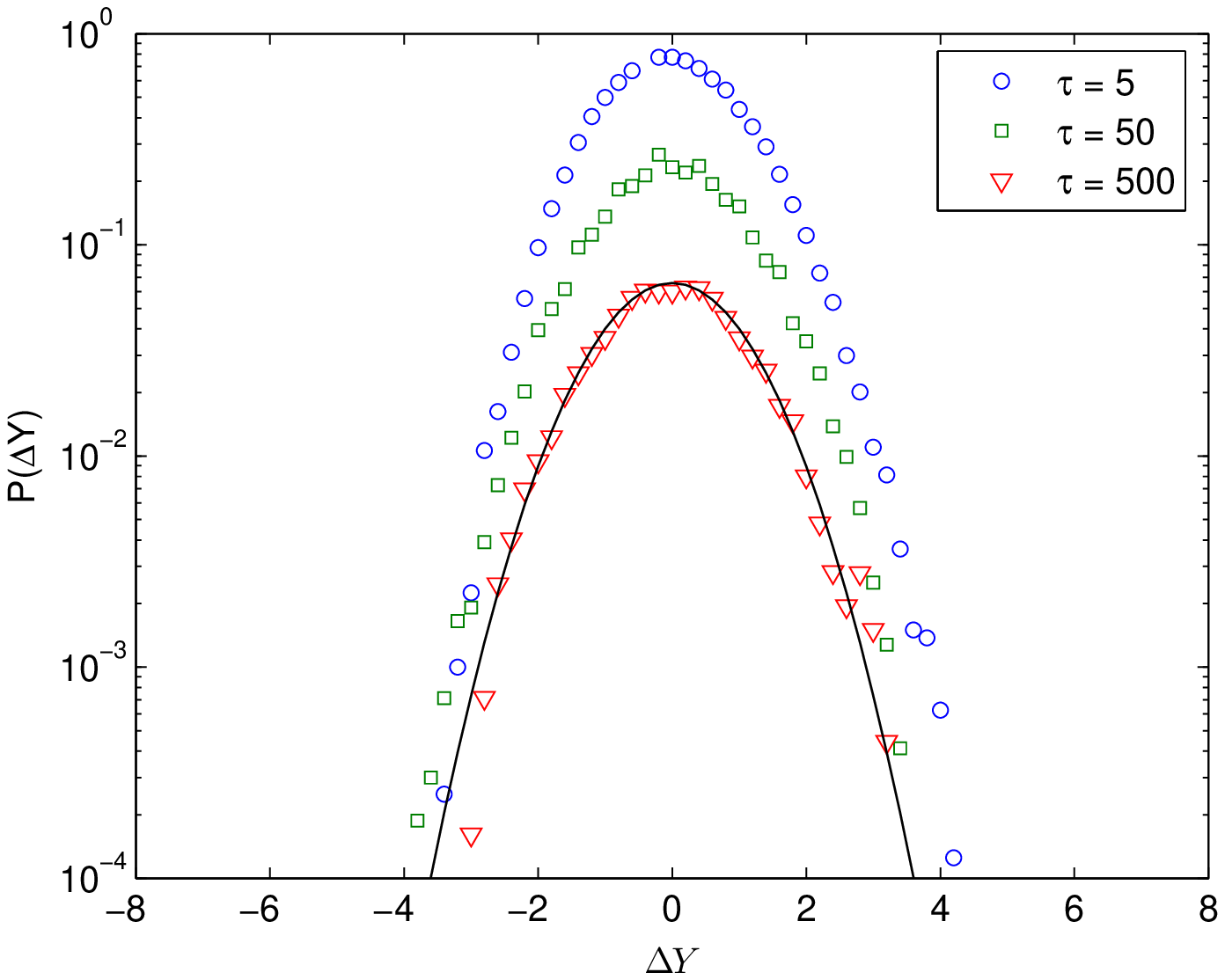}
 \caption{Probability distribution functions for two values of $\alpha$.
 Top: for $\alpha=0.05$ the PDFs for small scales are fat tailed.
 Bottom: for $\alpha=0.84$ the PDFs for all scales are gaussians.
 Shown as reference, the continuous line, the upside down parabola in both panels are gaussians
 PDFs with zero mean and unit standard deviation.}
 \label{pdf}
\end{figure}
In the Fig.~\ref{pdf} it is shown the PDFs for two values of
$\alpha$, for different dynamical regimes. In the bottom panel, for
$\alpha=0.84$, the two groups play independently from each other,
and the PDFs for all scales are gaussians. In the top panel, for
$\alpha=0.05$, the two groups play interactively and the PDFs are
fat tailed for small scales. This is a sign of intermittency
frequently found to be related to multifractal processes
\cite{frisch95}.

\section{Structure function analysis}
\label{sfa}

It is usual to characterize the MG dynamics only from the
statistical analysis point of view as was done in the section
\ref{stati}. In the following we introduce the Structure Function
(SF) analysis as a preliminary study to the multifractal scaling.

SF have been largely used in the study of turbulence \cite{frisch95}
and are defined in the following way \cite{sf}:
\begin{equation}
  S_q(\tau) \equiv \langle |Y(t+\tau) - Y(t)|^q \rangle \propto
  \tau^{q h(q)},
 \label{scaling}
\end{equation}
where $\langle~\rangle$ denotes the ensemble average. The SF can be
regarded as a generalization of the correlation functions (when
$q=2$). A signal that is scale invariant and self-similar is called
fractal when $h(q)$ is the same for all $q$, otherwise it is
multifractal \cite{paladin87}.
Another feature of the SF method is
its capability to identify nonstationarity in the data: for
stationary time series the exponent of the SF is zero, due to the
translational invariance of all statistics.

The time series analysis of the standard MG was done in
\cite{fff2003} using the unit root test and it was found that the MG
time series returns are stationary. To extend the analysis to the
present model, we first verified that the SF of the returns, $A(t)$,
for $q$ in the range $[1,6.5]$ were all flat for time series
$100,000$ points long, indicating that the signal is stationary.
Although the time series for the returns was found stationary in a
broad range, ones still needs to determine the scaling regime for
the cumulative sum of the returns, $Y(t)$, since its scaling range
does not necessarily follows the stationary range.

\begin{figure*}[!ht]
 \centering
 \includegraphics[width=\columnwidth]{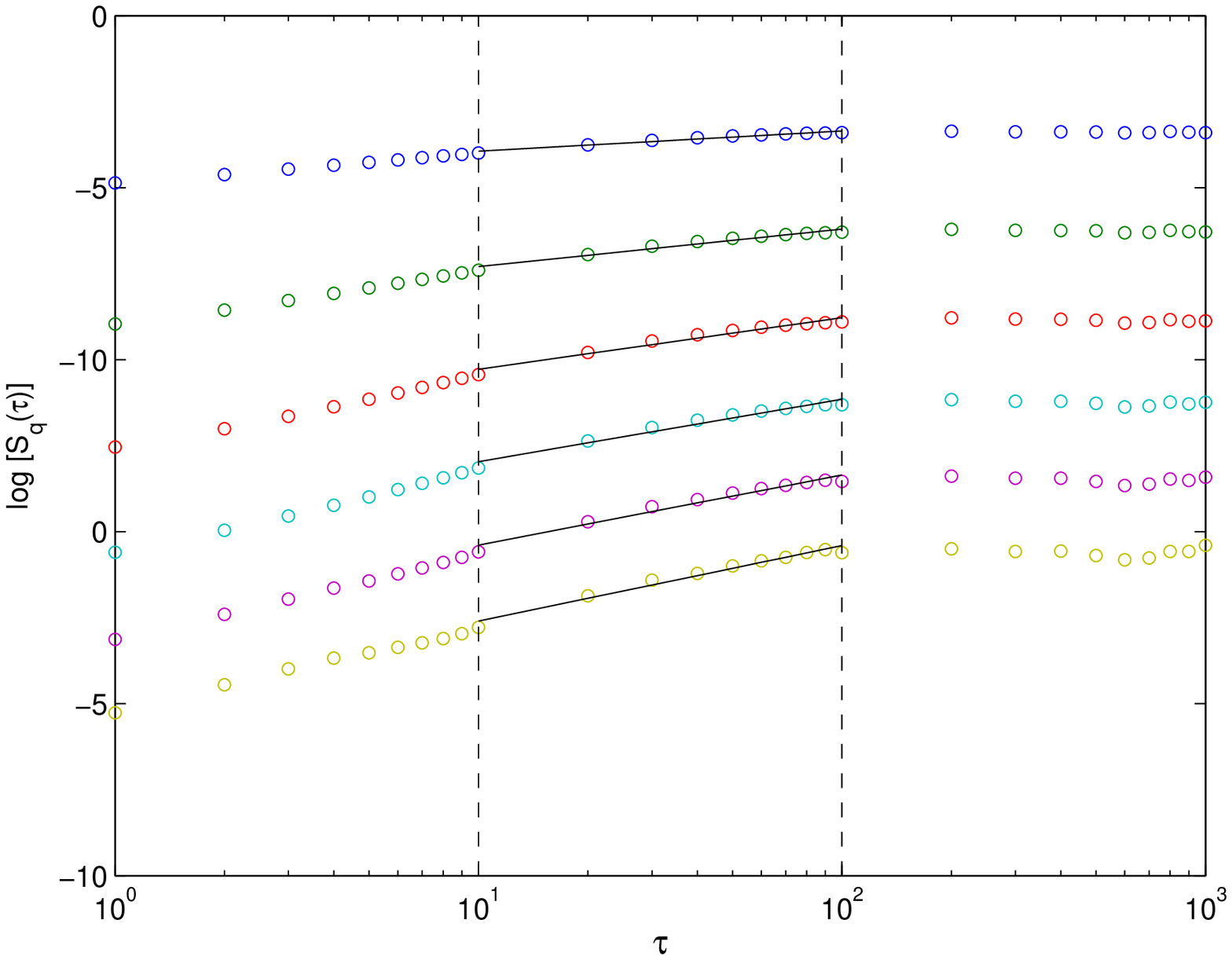}
 \includegraphics[width=\columnwidth]{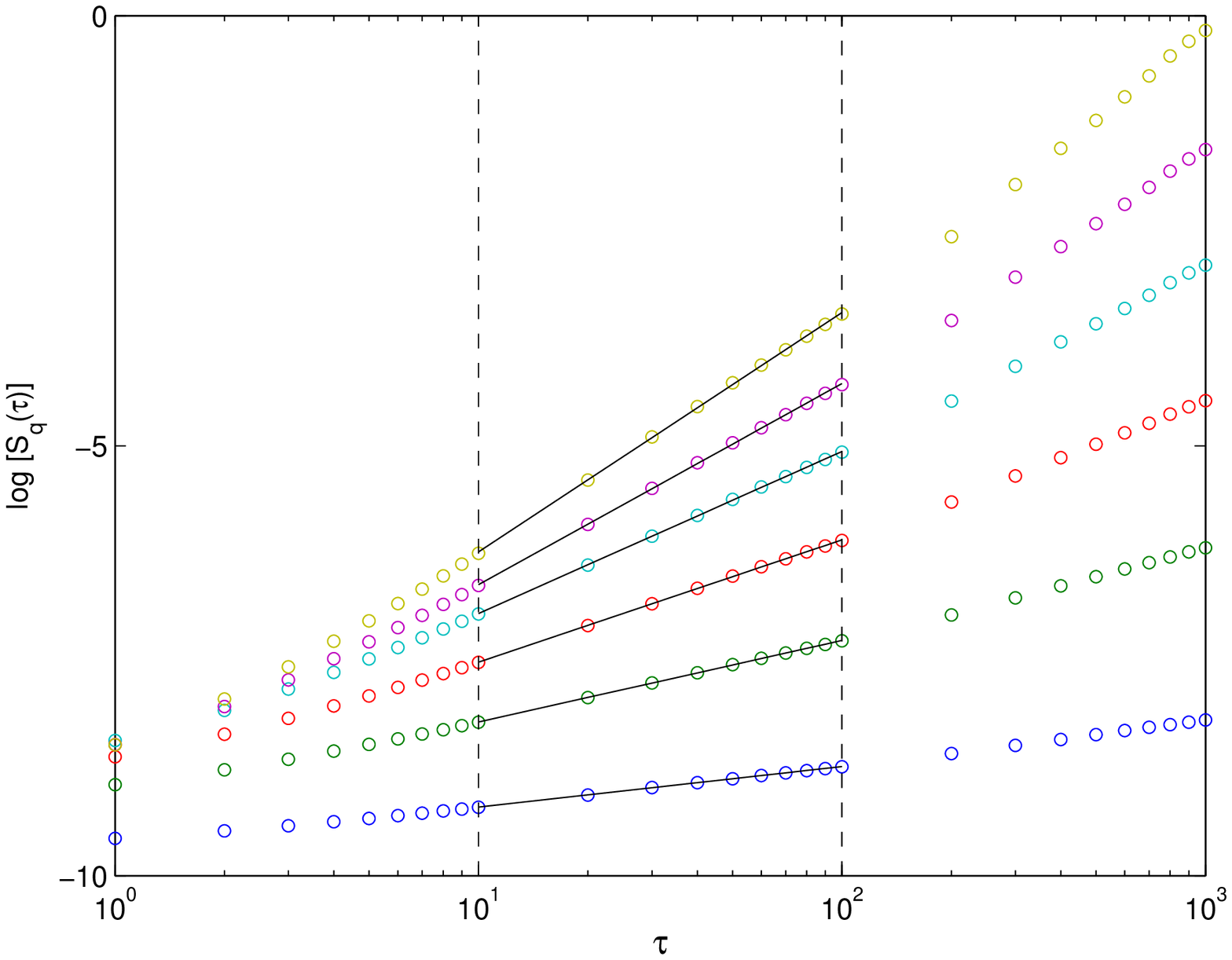}
 \includegraphics[width=\columnwidth]{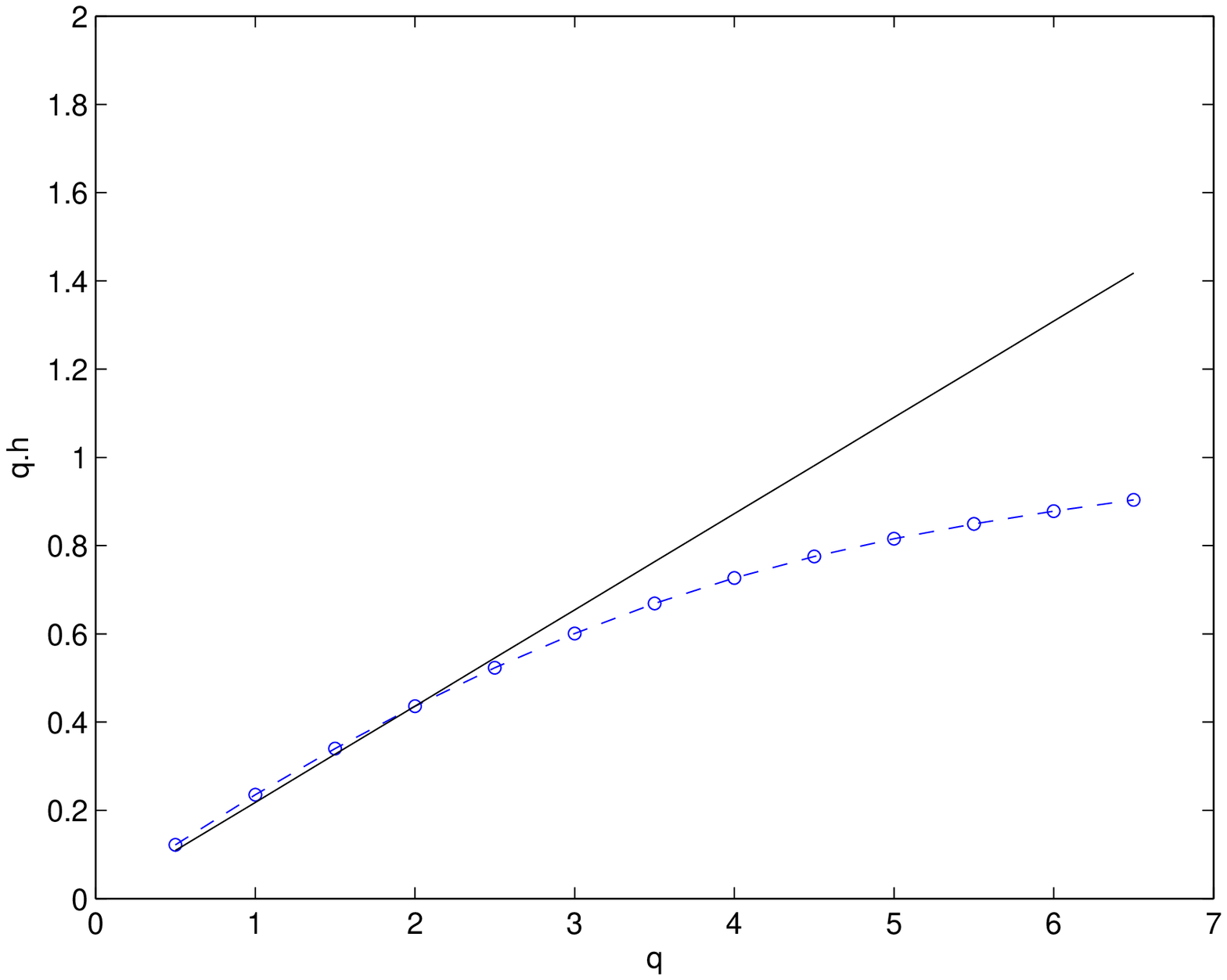}
 \includegraphics[width=\columnwidth]{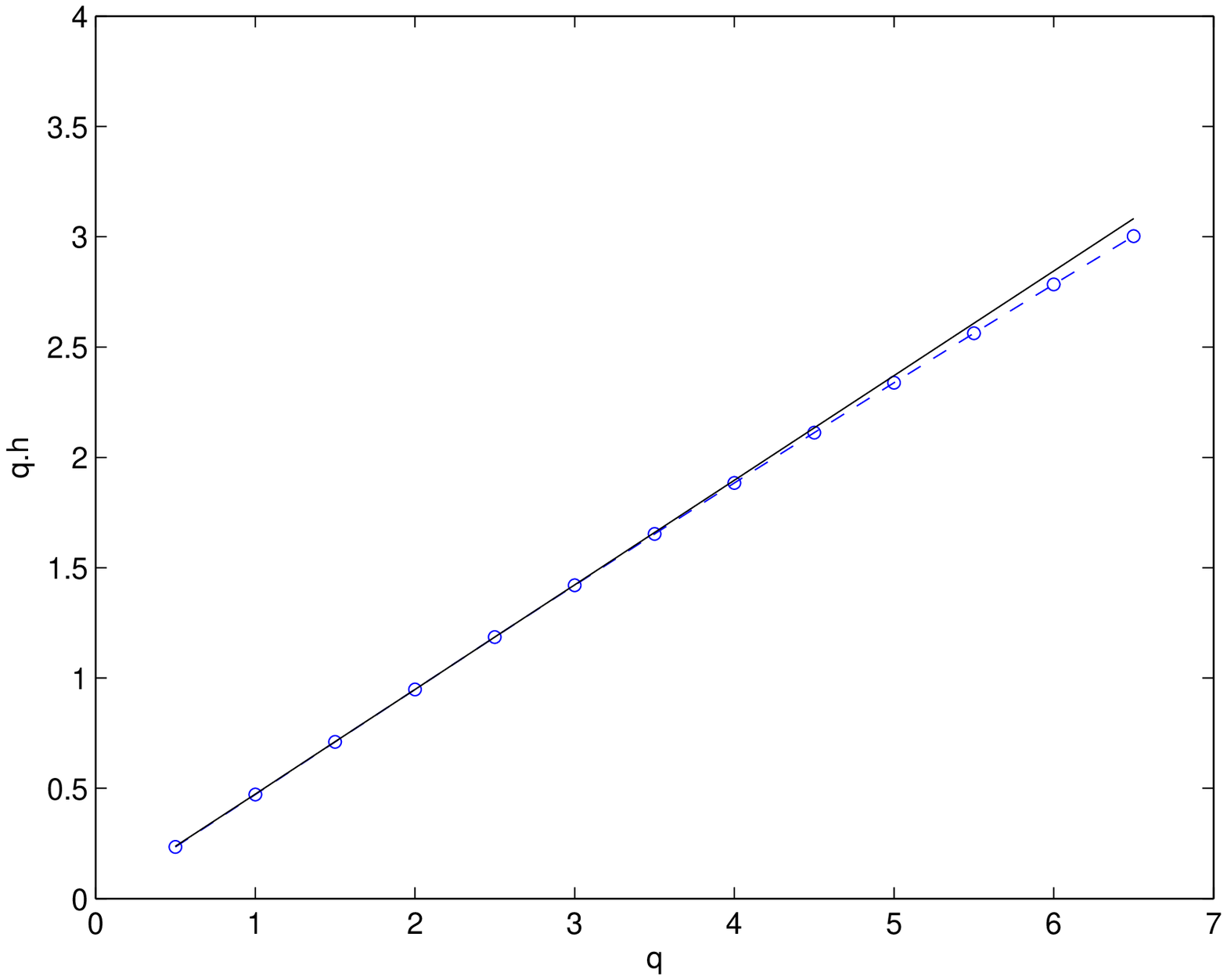}
  \caption{Top row: structure functions, $S_q(\tau)$, for the signal $Y(t)$,
  in arbitrary vertical scales, with dashed vertical lines indicating the
  scaling range. From top to bottom, the curves are for $q=1.0, 2.0, ... 6.0$.
  Bottom row: the scaling of the structure functions as given by Eq.~\ref{scaling}:
  when $h(q)$ is not constant with $q$, the signal is said multifractal
  (bottom, left panel), otherwise the signal is fractal (bottom, right panel).
  In these panels, the continous line slope gives $h(q=2)$, the Hurst exponent.
  The left column is for $\alpha=0.05$ and, the right one, for $\alpha=0.84$.
  }
\label{sffinal}
\end{figure*}

This difference between the stationary range and the scaling range
is seen in the SF of $Y(t)$ for $\alpha = 0.05$, panel a) in the
Fig.~\ref{sffinal}. These curves are clearly flat for $\tau > 1000$.
This does not occur for $\alpha = 0.84$, panel b) in the same
figure, that seems to have a scaling range broader than for $\alpha
= 0.05$. The superior limit for the scaling range is in general
different for different values of $\alpha$: we chose the value $\tau
= 100$, indicated by the dashed vertical line in the panels a) and
b), as the superior limit that is met for all values of $\alpha$.
While the scaling range limit on the right is believed to be due
information limits, on the left the scaling range is limited by a
different reason. Close to the interaction scale, $\tau = 5$ in the
present simulations, that is below the scaling range, the system
does not have yet statistical similarity. We will return to this
left limit in the next section. In short, panels a) and b) of
Fig.~\ref{sffinal} show clearly that the statistical $q$ order
moments follow scaling laws. Now we need to identify what is the
dependence of these scaling exponents with $q$.

The bottom line of Fig.~\ref{sffinal} shows the scaling of the SF as
given by Eq.~\ref{scaling}, $qh(q)$, against $q$. For $\alpha =
0.05$, panel c), the dependence is nonlinear with $q$, indicating
that the signal is multifractal, but for $\alpha = 0.84$, panel d),
the dependence is linear, indicating that the signal is fractal. For
reference, in the bottom panels, the slope of the continuous line
gives $H$, the Hurst exponent. Although $q$ may be in principle any
real number, negative moments are difficult to evaluate due to
divergence problems and will be treated using the WTMM method in the
next section.

\section{Multifractal analysis}
\label{multi}

A given signal can be considered as self-similar with scaling exponent
$h$ if its statistical properties are invariant under simultaneous
transformation of time $t$ and amplitude $Y(t)$ \cite{feder88},
\begin{equation}
  t \rightarrow b t\equiv t'; \hspace{2cm}  Y \rightarrow b^{h}Y\equiv Y';
 \label{selfaffine}
\end{equation}
where $b$ is an arbitrary positive constant and $h$ is a scaling exponent
given by the equation (\ref{scaling}).
A usual method to compute $h$ is based on the SF
approach, as shown in the section \ref{sfa}.

To obtain the full multifractal spectrum we make use of the Wavelet
Transform Modulus Maxima (WTMM) method. The wavelet family used in
this paper is the $n^{th}$-derivative of the Gaussian density
function(DOG$n$), whose wavelet transform has $n$ vanishing moments
and removes polynomial trends of order $n-1$ from the signal.
Because the scaling properties of the signal are preserved by the
wavelet transform, it is possible to obtain its multifractal
spectrum using this method. The number of vanishing moments for the
wavelet basis, $n$, is chosen to match the order up to ($n-1$) of
the polynomial trends in the signal.

The wavelet transform of a signal $Y(t)$ is defined as:
\begin{equation}
  T_{\psi}(\tau,b_0)=\frac{1}{\tau}\sum^{N}_{t=1} Y(t)\psi^*
  \Big(\frac{t-b_0}{\tau}\Big),
\end{equation}
where $\tau>0$ is the scale being analyzed, $\psi$ is the mother
wavelet DOG$n$ and $N$ is the number of points in $j$ direction. In
this paper we used $n=4$ for all analyses.

\begin{figure}[!ht]
  \centering
  \includegraphics[width=\linewidth,clip]{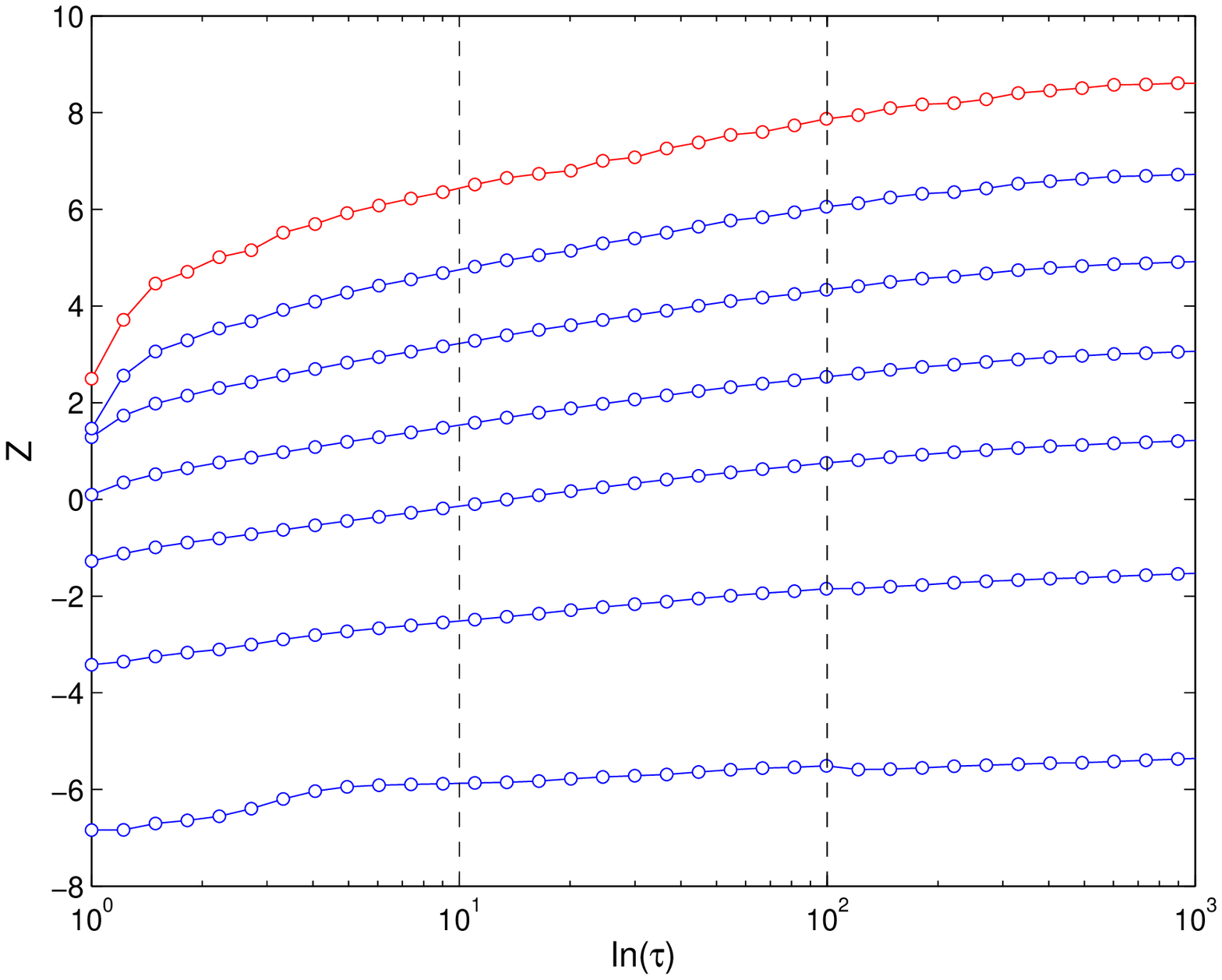}
  \includegraphics[width=\linewidth,clip]{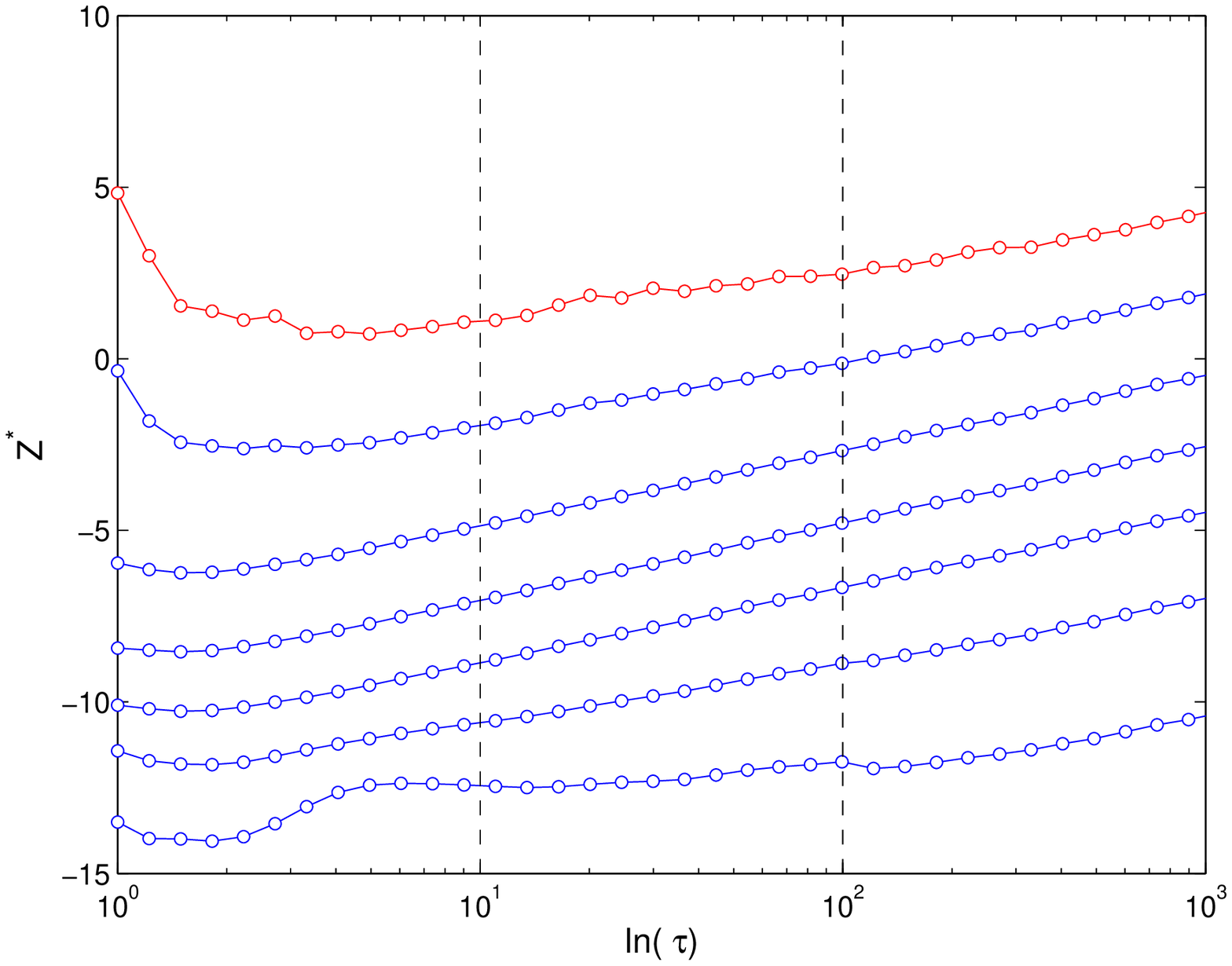}
  \caption{\label{legendre05} The functions $Z$ (top panel) and $Z^*$ (bottom panel),
  as given by Eq.~\ref{wtmm_a} and Eq.~\ref{wtmm_b}, for $\alpha=0.05$.
  The dashed vertical lines show the scaling region used to obtain $h(q)$.
  Each curve corresponds to different values of $q$.}
\end{figure}

\begin{figure}[!ht]
  \centering
  \includegraphics[width=\linewidth,clip]{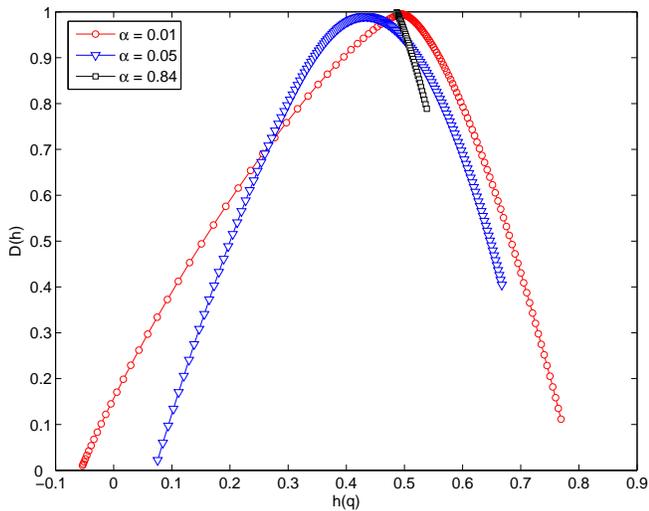}
  \caption{\label{mfs} Multifractal spectra for three values of $\alpha = 0.01, 0.05$
  and $0.84$. These values of $\alpha$ are, respectively, for a multifractal regime
  in the non-ergodic phase, a multifractal regime in the transition region and a fractal regime
  in the ergodic phase.}
\end{figure}

The statistical scaling properties of the singular measures found in
time series can be characterized by the singularity spectrum,
$D(h)$, of the H\"older exponents, $h$. A possible approach to
obtain the singularity spectrum directly from the time series is the
WTMM method \cite{muzy91, arneodo95}. The singularity spectrum,
$D(h)$, and the H\"older exponents, $h$, are obtained from the time
series by the following equations:
\begin{eqnarray}
  h(q)&=&\lim_{\tau \rightarrow 0}\frac{1}{\ln \tau}\sum_{\{ b_i(\tau) \}}
  \hat{T}_{\psi}[q;\tau,b_i(\tau)] \ln \left |T_{\psi}[\tau,b_i(\tau)] \right
  |\nonumber\\   &\equiv& \lim_{\tau \rightarrow 0}\frac{1}{\ln \tau} Z(q;\tau)
  \label{wtmm_a}
\end{eqnarray}
\begin{eqnarray}
 D(h)&=&\lim_{\tau \rightarrow 0}\frac{1}{\ln \tau}\sum_{\{ b_i(\tau) \}}
  \hat{T}_{\psi}[q;\tau,b_i(\tau)] \ln \big |\hat{T}_{\psi}[q;\tau,b_i(\tau)] \big |
  \nonumber\\ &\equiv& \lim_{\tau \rightarrow 0}\frac{1}{\ln \tau} Z^*(q;\tau)
  \label{wtmm_b}
\end{eqnarray}
where
\begin{equation}
  \hat{T}_{\psi}[q;\tau,b_i(\tau)] =  \frac{\left | T_{\psi}[\tau,b_i(\tau)] \right |^q}
  {\sum_{\{ b_i(\tau) \}} \left | T_{\psi}[\tau,b_i(\tau)] \right |^q}
\end{equation}
and the sum is over the set of the WT modulus maxima \cite{footnote}
at scale $\tau$, ${\{b_i(\tau) \}}$. The singularity spectrum,
$D(h)$, and the H\"older exponents, $h$, are obtained from scaling
range of the plots of Equations (\ref{wtmm_a}) and (\ref{wtmm_b}),
indicated by the dashed vertical lines in the Fig.~\ref{legendre05}.
Each curve on the figure corresponds to different values of $q$. For
large and small scales, the scaling regime is broken: to the right,
it saturates when the system reaches some physical limits and, to
the left, the system is in the range where the agents interacts. The
multifractal spectrum is in the Fig.~\ref{mfs}.

One way to interpret the multifractal spectrum in a physical sense
is by comparison with the Hurst exponent for known signals, for
instance, the \textit{fractional Brownian motion}
\cite{feder88,mandelbrot68}. The fractional Brownian motion can be
classified according to the probabilities of its fluctuations. The
usual Brownian motion, obtained from a Gaussian white noise, has the
same probability of having positive or negative fluctuations and
$H=0.5$. A fractional Brownian motion with $H < 0.5$ is more likely
to have the next fluctuation with opposite sign with respect to last
one -- it is said to be antipersistent. Conversely, a fractional
Brownian motion with $H > 0.5$ is more likely to have the next
fluctuation with the same sign as the last one -- it is said to be
persistent. Antipersistent signals have more local fluctuations and
seem more irregular in small scales. Their variance diverges slower
with time than the variance of persistent signals. Such signals
fluctuate on larger scales and look smoother. This discussion is
done in \cite{superficie2006} and a similar, but more detailed
interpretation, is given in \cite{whatcolor2000}.

Now we will analyse for the fractal/multifractal properties of the
models' dynamics as a function of the control parameter $\alpha$. In
the ergodic regime, $\alpha=0.84$, Fig. \ref{mfs} shows that the
multifractal spectrum is narrow, almost collapsed to a single point.
As already expected, this region is clearly fractal, since the time
series for the returns is Gaussian. Just before the transition, in
the subcritical regime, the spectrum is wide, a sign of
multifractality.

The multifractal spectrum may be represented by its extrema points,
i.e., its minima on the left, $h_l$, and on the right, $h_r$, as
well as the maximum  (top), $h_0$. It is worth to note that a wide
spectrum (difference between $h_r$ and $h_l$) is a clear evidence
for the multifractal character of the time series.

\begin{figure}[!ht]
  \centering
  \includegraphics[width=\linewidth,clip]{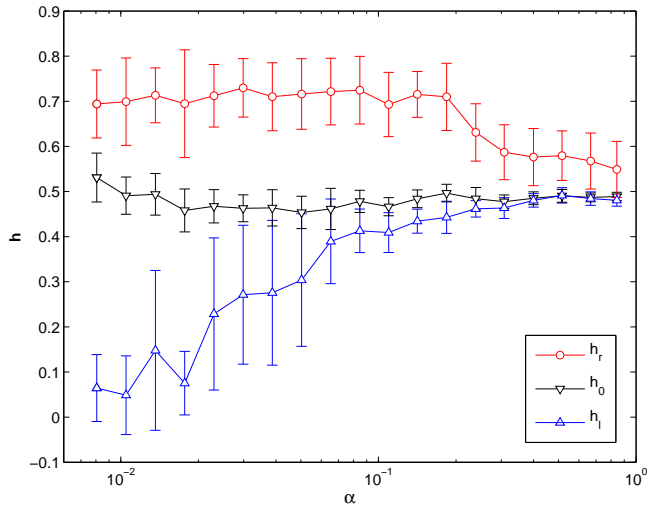}
  \caption{\label{pn} Extrema points of the Multifractal spectra, $h_l$, $h_0$ and $h_r$,
  for $\alpha$ values from $0.008$ to $0.84$.
  The signature of multifractality is in the fact that
  the difference $h_r-h_l$ increases as $\alpha$ decreases below $\alpha_c=0.2$.
  In the region $\alpha>\alpha_c$, where data are Gaussian, this difference should be vanishingly
  small for time series with infinite number of points.
  The error bars were obtained from 20 realizations of different initial conditions.}
\end{figure}

We studied the behavior of the extrema points as a function of
$\alpha$ to characterize the transition from the fractal to the
multifractal regime, as shown in the Fig.~\ref{pn}. This figure
shows that the spectrum width increases as $\alpha$ decreases, going
from a fractal to multifractal regime. This transition is smooth and
exhibits large fluctuation of $h_l$ in the nonergodic phase. The
point $h_l$, which shows the scaling of the fluctuation of the
signal, is sensitive to the realization of the trajectory. On the
other hand, $h_r$ captures the scaling of the small fluctuation. The
point $h_0$ fluctuates slightly close to 0.5 which is the value
observed for random walks.

In the nonergodic, sub-critical phase, the system is said to represent an efficient
market, where no arbitrage is possible: this is coherent with the multifractal nature
of the time series since this dynamics is richer than the fractal one, that is typical
on the ergodic phase.

\section{Conclusion}
\label{final}

We showed that the MG model presents a very rich dynamics with
anomalous fluctuations that arise due to strong correlation similar
to the one observed in systems driven out of equilibrium. One of the
more difficult stylized fact to obtain from models for financial
systems are the long range correlations that generate the fractal
and multifractal properties present in real time series. The
synchronous MG model  does not present multifractality. However,
only one ingredient added to the model generates multifractal region
on the space control parameter $\alpha$. This ingredient was the
broken of synchronicity. As the statistical properties of the MG are
preserved, the observed multifractal regime belongs to the
nonergodic phase, where the market is informationally efficient.

Although the MG time series are not stationary, their increments
are. Using the SF approach we detected the stationary range of the
increments and the scaling range from the time series. From the
linear (non-linear) behavior of the SF we identified the fractal
(multifractal) regimes as functions of $\alpha$. To look at the
negative values of the moments $q$ we used the WTMM and obtained the
full multifractal spectrum against $\alpha$.

\acknowledgments

FFF thanks the Conselho Nacional de Desenvolvimento Cient\'{\i}fico
e Tecnol\'ogico (CNPq) for financial support and Instituto de
F\'{\i}sica Te\'orica for hospitality.

\end{document}